# Imprinting control regions (ICRs) are marked by mono-allelic bivalent chromatin when transcriptionally inactive

Stéphanie Maupetit-Méhouas[1,2,3,†], Bertille Montibus[1,2,3,†], David Nury[1,2,3,†], Chiharu Tayama[4], Michel Wassef[5], Satya K. Kota[6], Anne Fogli[1,2,3], Fabiana Cerqueira Campos[1,2,3], Kenichiro Hata[4], Robert Feil[6], Raphael Margueron[5], Kazuhiko Nakabayashi[4], Franck Court[1,2,3] and Philippe Arnaud[1,2,3,*]

[1]CNRS, UMR6293, F-63001 Clermont-Ferrand, France, [2]Inserm, U1103, 63001 Clermont-Ferrand, France, [3]Université Clermont Auvergne, Laboratoire GReD, BP 10448, 63000 Clermont-Ferrand, France, [4]Department of Maternal-Fetal Biology, National Research Institute for Child Health and Development, 2-10-1 Okura, Setagaya, Tokyo 157-8535, Japan, [5]Institut Curie, 26 Rue d'Ulm, 75005 Paris, France; INSERM U934, 26 Rue d'Ulm, 75005 Paris, France; CNRS UMR3215, 26 Rue d'Ulm, 75005 Paris, France and [6]Institute of Molecular Genetics, CNRS UMR-5535 and University of Montpellier, 1919 route de Mende, 34293 Montpellier, France

## ABSTRACT

**Parental allele-specific expression of imprinted genes is mediated by imprinting control regions (ICRs) that are constitutively marked by DNA methylation imprints on the maternal or paternal allele. Mono-allelic DNA methylation is strictly required for the process of imprinting and has to be faithfully maintained during the entire life-span. While the regulation of DNA methylation itself is well understood, the mechanisms whereby the opposite allele remains unmethylated are unclear. Here, we show that in the mouse, at maternally methylated ICRs, the paternal allele, which is constitutively associated with H3K4me2/3, is marked by default by H3K27me3 when these ICRs are transcriptionally inactive, leading to the formation of a bivalent chromatin signature. Our data suggest that at ICRs, chromatin bivalency has a protective role by ensuring that DNA on the paternal allele remains unmethylated and protected against spurious and unscheduled gene expression. Moreover, they provide the proof of concept that, beside pluripotent cells, chromatin bivalency is the default state of transcriptionally inactive CpG island promoters, regardless of the developmental stage, thereby contributing to protect cell identity.**

## INTRODUCTION

Genomic imprinting is a specialized mechanism of transcriptional regulation whereby a subset of mammalian genes is expressed only from one allele, according to its parental origin. Most of the about 150 imprinted genes identified to date are involved in key biological processes, such as cell proliferation, fetal and placental growth, metabolic adaptation, as well as neurological processes and behavior. Consequently, genomic imprinting must be correctly regulated during the entire life-span (1) and its misregulation is causally involved in several growth and behavioral syndromes in humans (2,3).

Genomic imprinting regulation relies on DNA methylation. Notably, all the discrete CpG-rich cis-acting regions that control imprinted expression (imprinting control regions, ICRs) overlap with a differentially methylated region (DMR) that harbors allelic DNA methylation inherited from the male or female gamete (paternal and maternal germline DMRs) and subsequently maintained throughout development (4). This allelic DNA methylation can be 'read' in different ways to ensure the appropriate tissue- and developmental-specific allele-specific expression of entire clusters of imprinted genes (5). Based on this constant observation, and for clarity, DMRs that acquire differential DNA methylation in the germline will be called ICRs thereafter.

Therefore, the allelic DNA methylation signature at ICRs is the first and necessary level of regulation that must be tightly controlled. Specifically, besides being targeted by

*To whom correspondence should be addressed. Tel: +33 473178380; Fax: +33 473276132; Email: philippe.arnaud@udamail.fr
†These authors contributed equally to this work.
Present address: Satya K. Kota, Harvard School of Dental Medicine and Harvard Medical School, Boston, USA.

DNA methylation in a germline-specific manner, a specific feature of ICRs is their ability to maintain an allelic-specific methylation pattern during somatic development, including during the peri-implantation reprogramming steps (6).

In the last years, most work focused on the identification of the mechanisms whereby DNA methylation is brought and maintained specifically to one of the two parental alleles. In the emerging model, transcriptional read-through events and removal, or absence of the permissive H3K4me chromatin mark act in a concerted manner to recruit the *de novo* methyltransferase DNMT3A and its non-catalytic co-factor DNMT3L to ICRs in the germline (7–9). The resulting DNA methylation resists to the pre-implantation demethylation wave thanks to the presence of the maternal protein PGC7/Stella and the zing finger protein ZFP57 (10,11). Specifically, ZFP57 binding discriminates ICRs from the thousands of other CpG islands that are methylated in the germline and directly contributes to the selective protection of imprinted methylation at ICRs against the genome-wide pre-implantation demethylation wave (12). After implantation, the DNA methyltransferase DNMT1 ensures the faithful maintenance of DNA methylation during each cell cycle (13,14).

Conversely, the mechanisms whereby the opposite parental allele remains unmethylated are less documented. The presence of H3K4me, which is anti-correlated with DNA methylation in mammalian genomes (15–17), might be instrumental. Indeed, the DNMT3 family interacts with histone H3 only when it is unmethylated on lysine 4, suggesting that genomic sequences enriched for H3K4me cannot recruit the *de novo* DNA methylation machinery (9,18,19). This is consistent with the observation that H3K4me2 belongs to the canonical chromatin signature of ICRs (20). H3K4me protective role is further supported by the observation that in pro-spermatogonia, when paternal ICRs acquire DNA methylation, maternal ICRs are active promoters and are enriched for H3K4me3 (21,22). RNA-DNA hybrids (R-loops) formed at active promoter regions, as observed at the human SNRPN and murine *Airn* ICRs, could also contribute to the protection against the DNA methylation machinery (23). The finding that single-stranded DNA efficiently recruits SET-domain H3K4 methyltransferases (24) raises the possibility that R-loop formation contributes to H3K4me3 accumulation. Thus, at ICRs that are active promoters, the transcriptionally active allele is H3K4me-enriched and protected from DNA methylation. Nonetheless, their promoter activity cannot explain on its own the ICR allele-specific unmethylated status because this activity is often tissue- and/or developmental-stage specific. This feature highlights the dual constrains acting on the unmethylated allele of ICRs. Indeed, it must be constitutively unmethylated even when transcriptionally inactive. This suggests that a dedicated fine-tuned regulation mechanism is required and bivalent chromatin emerges as a relevant candidate.

Bivalent chromatin domains are unusual because histone H3 can concomitantly be marked by the 'active' H3K4me and the 'repressive' H3K27me3 modifications. Bivalent domains were initially detected at promoters of many genes in both human and mouse ES cells (15,25,26). In the proposed model, bivalent chromatin domains repress gene transcription through H3K27me3, while keeping genes 'poised' for alternative fates induced by specific developmental cues. (25). Consistently, the resolution of these bivalent domains into either H3K4me3 or H3K27me3, as observed upon stem cell differentiation, is believed to stably mark these regions for activation or repression, respectively (16).

However, their precise role in development remains controversial because probing the function of bivalent domains in developing organisms remains a challenge (27).

Bivalency at imprinted loci has been reported at the maternally-methylated *Grb10* ICR and at the paternally-methylated non-promoter *H19* ICR (28–31). We previously showed that the bivalent domain on the unmethylated allele of the *Grb10* ICR, which is conserved in mice and humans, contributes to the control of its paternal brain-specific expression (28,29). Moreover, an *in silico* study suggested the implication of chromatin bivalency in imprinting, although it was conducted without taking into account the parental origin of the analyzed histone marks and was limited to three cell types (32).

Here, we wanted to determine whether mono-allelic chromatin bivalency is an ICR common signature that could contribute to the fine-tuned regulation of imprinted gene expression. To this purpose, we conducted a comprehensive analysis of maternally imprinted ICRs (mat-ICRs), which, unlike paternally imprinted ICRs, are all associated with promoter regions, to determine their chromatin signature relative to their transcriptional activity in different tissues. Our main observation is that chromatin bivalency is the default state of the unmethylated allele in non-expressing tissues. Our data further support a structural role for chromatin bivalency at ICRs by maintaining their integrity during somatic development.

## MATERIALS AND METHODS

### Material collection

Material was obtained from reciprocal crosses between C57BL/6J (B6) with *Mus musculus molossinus* JF1/Ms mice, (B6xJF1) F1 and (JF1xB6) F1, respectively, referred to as 'BJ' and 'JB' in the text. Whole embryos, placenta and isolated brain and liver tissues were recovered at various developmental stages, as indicated in the text. MEFs were derived from E13.5 carcasses.

*Dnmt3L*$^{-/+}$ mouse embryos were obtained by crossing homozygous *Dnmt3L*$^{-/-}$ females (129SvJae-C57BL/6 hybrid genetic background) with WT JF1 male mice (*M. musculus molossinus*). E9.5 *Dnmt3L*$^{-/+}$ conceptuses were removed from pregnant mothers. Tail DNA was used for genotype analysis by PCR as described previously (33).

### ES and iMEF cell lines

The derivation and characterization of the reciprocal hybrid ES cell lines between the *M. m. domesticus* strain C57BL/6J and the *M. m. molossinus* strain JF1 (lines BJ1 and JB1, respectively) are described in Kota *et al.* (39). These ES cell lines are maintained on gelatin-coated dishes in ESGRO-complete-plus medium (Millipore, SF001-500P) that contains LIF and BMP4.

DNA and RNA samples from the B1.3 *Eed*$^{-/-}$ and WT ES cell lines (26) were from Amanda Fisher's laboratory (Lymphocyte Development Group, MRC, London, UK).

MEF cells were isolated from E13.5 *Ezh2* flox/flox;ROSA26-CreERT2 mouse embryos and then infected with the pBABE-hygro p53-DD retroviral construct (addgene 9058) to generate p53-DN immortalized MEFs (iMEFs). iMEFs were grown in Dulbecco's modified Eagle's medium (Gibco, 11960-044) with 10% fetal bovine serum (PAA, A15-101), 2 mM glutamine (PAA M11-006) and 1% MEM Non-essential amino acids solution (Gibco 11140-035) at 37°C in 5% CO2. For conditional *Ezh2* deletion, iMEFs were incubated with 4-hydroxytamoxifen (Sigma, T176) at a final concentration of 1 μM.

### JF1/Ms genome sequencing and SNP identification and validation

The JF1/Ms strain was obtained from the Mammalian Genetics Laboratory at the National Institute for Genetics (Mishima, Shizuoka, Japan). Genomic DNA was extracted from the liver of a 43-week-old female using the DNeasy Blood and Tissue kit (Qiagen, 69504). The genomic DNA library was prepared using the TruSeq DNA Sample preparation kit (Illumina, FC-121–2001). The library (average fragment size 437 bp) was sequenced using the TruSeq SBS Kit v3 reagent (Illumina, FC-401-3001) and the HiSeq1000 platform at the National Center for Child Health and Development. The sequences (280 Gb in total) obtained by 100-bp paired-end sequencing were deposited at the DNA Data Bank of Japan (DDBJ) under the DDBJ Sequence Read Archive (DRA) accession number DRX005582 (http://trace.ddbj.nig.ac.jp/dra/index_e.html). Sequences were aligned to the mouse reference genome (mm10) using the BWA software (http://bio-bwa.sourceforge.net/).

For each region of interest, single nucleotide variations, insertions and deletions in the JF1/Ms genome, compared to the reference C57BL/6J genome, were detected using the Genome Analysis Toolkit (GATK) 1.5 (https://www.broadinstitute.org/gatk/). Identified SNPs were validated by direct sequencing of PCR products of the regions of interest obtained respectively from C57BL/6J, JF1 and C57BL/6J/JF1 liver DNA. Details of the SNPs used in this study are given in Supplementary Table S1.

### DNA extraction and bisulfite sequencing

DNA extraction was done as previously described (34). Bisulfite conversion was performed by using the EZ DNA methylation$^{TM}$ Gold Kit from Zymo (ref. D5006), according to the manufacturer's instruction. PCR amplifications, cloning and sequencing were performed as previously described (34). Details on the primers used are in Supplementary Table S1.

### RNA extraction and expression analysis

Both commercial and on-site extracted RNA samples were used in this study. To circumvent potential inter-individual variations, quantitative gene expression analyses were performed using a commercial panel of total RNA (mouse total RNA master panel; Ozyme 636644) obtained from pooled samples isolated from several hundreds of mouse embryos and adults. For analysis of BJ and JB material, total RNA was extracted using the Trizol Reagent (Life Technologies, 15596018). After digestion with RNase-free DNase I (Life Technologies, 180868-015), first strand cDNA was generated by reverse transcription with Superscript-III (Life Technologies, 18080085) using randomized primers and 1 μg of RNA. Duplicate sets of samples were produced without reverse transcriptase to detect amplification from contaminating DNA.

*Allelic analysis.* For each locus of interest, the parental-allele origin of expression was assigned following direct sequencing of the cognate RT-PCR product that encompassed a strain-specific SNP (SNP details in Supplementary Table S1).

*Microfluidic-based quantitative analysis.* First-strand cDNA was pre-amplified for 14 cycles with the pool of primers used for the RT-qPCR analysis and the TaqMan PreAmplification Master Mix (Life Technologies, 4488593). RT-qPCRs were then performed and validated on Fluidigm 96.96 Dynamic Arrays using the Biomark HD system (Fluidigm Corp.) according to the manufacturer's instructions.

The relative gene expression was quantified using the $2^{-\Delta\Delta Ct}$ method (35) that gives the fold changes in gene expression normalized to the geometrical mean of the expression of the housekeeping genes *Arbp*, *Gapdh*, *Tbp* and, according to the analysis, relative to one calibrator, as indicated in the text or figure legend. For each condition, the presented data were obtained from two independent experiments, each analyzed in duplicate.

*RT-qPCR analyses.* RNA was reverse transcribed as described before and amplified by real-time PCR with a SYBR Green mixture (Roche) using a LightCycler® 480II (Roche) apparatus. For each condition, analyses were repeated four times, each in duplicate. The primer sequences are in Supplementary Tables S1 and S2.

### Chromatin immuno-precipitation

ChIP of native chromatin was carried out as described in Wagschal *et al.* (36). Results presented in this article were obtained from at least three and up to seven ChIP assays performed using independent chromatin preparations, as indicated in the figure legends. Details of the used antisera are in Supplementary Table S3. Sequential ChIP was performed as described by Bernstein *et al.* (25), using cross-linked chromatin obtained from BJ1 ES cells.

### Analysis of immunoprecipitated chromatin

*Allelic analysis.* In the input and antibody-bound fractions for each antiserum used, the parental alleles were differentiated by direct sequencing of the PCR products encompassing a strain-specific SNP in the regions of interest (SNP details in Supplementary Table S1). For PCR amplification (30–35 cycles, depending on the analyzed region), the HotStar Taq DNA polymerase (Qiagen, 203205) was used, according to the manufacturer's recommendations. The relative allelic ratios were quantified from sequence files in ABI format using the Mutation quantification module of the Mutation Surveyor® DNA variant analysis software (Softgenetics, http://www.softgenetics.com/mutationSurveyor.html).

*Quantitative analysis.* Input and antibody-bound fractions were quantified by real-time PCR amplification with a SYBR Green mixture (Roche) using a LightCycler® 480II (Roche) instrument. Background precipitation levels were determined by performing mock precipitations with a non-specific IgG antiserum (Sigma C-2288) and were only a fraction of the precipitation levels obtained with specific antisera. Bound/ input ratios were calculated and were normalized, according to the antiserum used, against the precipitation level at the *Rpl30*, *HoxA3* and *HoxD8* promoters or IAP elements, as indicated in the figure legends. The primers used are in Supplementary Table S2.

### *In silico* expression analysis in *Ezh2*$^{-/-}$ somatic cells

Raw data were extracted from publicly available RNA-seq data on E12.5 mouse cardiomyocytes (GSE 29997; 37) and resting and activated spleen T regs (GSE58998; 38) and plotted using R.

## RESULTS

### A subset of mat-ICRs is marked by mono-allelic bivalent chromatin in ES cells

We first determined whether, besides the *Grb10* ICR, other mat-ICRs harbor mono-allelic bivalent chromatin in mouse ES cells. To this aim, we took advantage of the recently derived reciprocal hybrid ES lines between the *M. m. domesticus* strain C57BL/6J and the *M. m. molossinus* strain JF1 (39), named BJ1 and JB1 cells, respectively. Bisulfite analysis showed that mono-allelic, maternal DNA methylation was faithfully maintained at the ICRs of these cells in our culture conditions (Figure 1A; Supplementary Figure S1). We then performed immunoprecipitation of native chromatin (ChIP) to determine the histone modification pattern at 14 of the 20 characterized mat-ICRs for which we had identified SNP polymorphisms between the C57BL/6J and JF1 strains (Supplementary Table S1). Using these polymorphisms, we analyzed the allelic enrichment of H3K4me2, H3K4me3, H3K27me3 and H3K9me3. At all tested ICRs, the unmethylated paternal allele was marked by H3K4me2 and H3K4me3, while the methylated maternal allele was associated with H3K9me3 (Figure 1A, Supplementary Figure S1). Beside this canonical histone modification signature (reviewed in (40)), we observed that H3K27me3, which marks most but not all ICRs on the methylated allele, was also present at the unmethylated allele of seven ICRs (*Inpp5f-v2*, *Plagl1*, *Nnat*, *GnasXL*, *Peg10*, *Nap1L5* and *Igf2r*). Its association with the permissive marks H3K4me2 and H3K4me3 is reminiscent of the bivalent chromatin signature (Figure 1A, Supplementary Figure S1).

After ChIP with an anti-H3K27me3 antibody, we confirmed by bisulfite sequencing the presence of both the maternally methylated and the paternally unmethylated allele of these seven ICRs in the precipitated chromatin (Supplementary Figure S2). Thus, differently from most CpG islands (41,42), DNA methylation and H3K27me3 can coexist at a subset of ICRs, as already shown in other studies (*e.g*: 43–47).

Next, we observed that, unlike the other mat-ICRs, transcripts initiating from ICRs with bivalent chromatin on the paternal allele (see Supplementary Table S1 for details) were barely or not detectable in ES cells, suggesting a repressive structure that is homogeneously present in all cells (Figure 1B). This was further supported by *in silico* analysis of the deposition of total RNA PolII and H3K64ac, a histone mark associated with active promoters (48). Strikingly, we observed that H3K64ac discriminated the mat-ICRs associated with bivalent chromatin from the transcriptionally active mat-ICRs (Supplementary Figure S3). A similar trend is observed for total RNA PolII, that is nonetheless enriched at *Nnat* and *Inpp5f-v2* ICRs suggesting a primed configuration at these loci, a signature found, genome wide, at 2/3 of bivalent domains (49) (Supplementary Figure S3). Conversely, the *Peg10* locus, which showed a sustained transcription level (Figure 1B), was consistently enriched for H3K64ac (Supplementary Figure S3). Nonetheless, sequential ChIP approaches, in which two successive rounds of immunoprecipitation were performed using antibodies against H3K4me3 and then against H3K27me3 or reciprocally, confirmed that these seven mat-ICRs, including the *Peg10* ICR, carried both histone modifications on their paternal unmethylated allele (Supplementary Figure S3). This suggests that transcription occurs at *Peg10* despite the presence of H3K27me3.

Combined, our observations reveal that, including *Grb10*, eight mat-ICRs out of 15 studied have bivalent chromatin on their unmethylated paternal allele in ES cells.

### Mat-ICRs associated with bivalent chromatin in ES cells comprise tissue-specific promoters

To determine whether this bivalent chromatin signature could be involved in the control of tissue-specific gene expression, we profiled, using a microfluidic quantitative RT-PCR approach, the expression pattern of transcripts initiating from these eight mat-ICRs regions at different stages of mouse development and in different adult tissues. To ensure that the possible inter-individual variation in expression level did not interfere with the analysis, each RNA sample was obtained from pooled samples isolated from several hundred animals. We also included RNA obtained from mouse embryonic fibroblasts (MEFs) derived from C57BL/6J x JF1 (BJ) E13.5 embryos.

Transcripts initiating from these eight regions were strongly expressed only in few tissues (Figure 2). Specifically, expression of *Inpp5f-v2*, *Nap1l5* and *Grb10* (isoform initiating from the ICR and referred as 'Grb10 pat-isoform') was almost exclusively restricted to neural tissues (spinal cord and brain), in agreement with the literature (e.g. 28,50,51). In addition to expression in neural tissue, *Nnat*, *Plagl1* and *GnasXL* were also strongly expressed in E11 to E17 embryos, while *Peg10* was mostly detected in placenta. On the other hand, *Airn* that is not expressed in a subset of tissues was homogenously expressed, albeit at low level, in several samples. These results show that at the eight mat-ICRs associated with bivalent chromatin, transcripts are expressed in a tissue- and developmental-specific fashion.

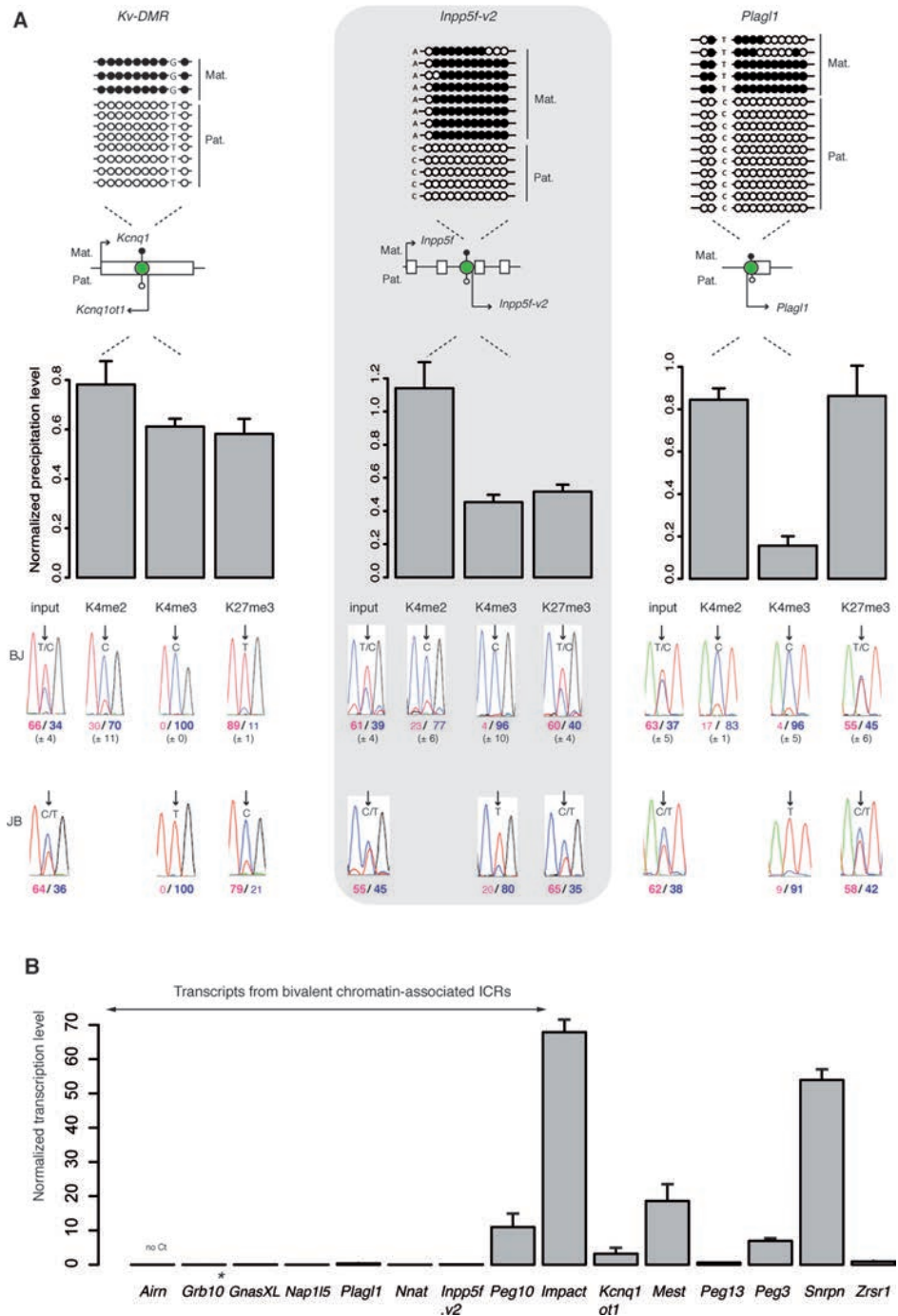

**Figure 1.** Bivalent chromatin marks the unmethylated allele of a subset of maternal ICRs in ES cells. (**A**) Schematic representation of the analyzed ICRs (green circles) and their position relative to the main associated gene(s). The methylation patterns are symbolized by lollipops (black: methylated; white: unmethylated). The name of the transcripts initiating from the ICR region are also indicated. *Kv-DMR* is an example of an ICR with a canonical chromatin signature. The *Inpp5f-v2* and *Plagl1* DMRs are examples of ICRs marked by mono-allelic bivalent chromatin. The upper panels show representative bisulfite-based sequencing data obtained from BJ1 ES cells. Each horizontal row of circles represents the CpG dinucleotides on an individual chromosome. Solid circles, methylated CpG dinucleotides; open circles, unmethylated CpG dinucleotides. The parental origin (Mat., maternal; Pat., paternal) was determined using the indicated strain-specific SNPs. The lower panels show the data obtained by native ChIP using anti-H3-K4me2, -K4me3 and -K27me3 antibodies. These three histone marks were similarly enriched at the analyzed ICRs in BJ1 ($n = 5$) and JB1 ($n = 1$) ES cells. The precipitation level was normalized to that obtained at the *Rpl30* promoter (for H3K4me2 and H3K4me3) and at the *Hoxa3* promoter (for H3K27me3). The allelic distribution of each histone mark was determined by direct sequencing of the PCR product encompassing a strain-specific SNP in the analyzed region. The mean values (± standard deviation) of the relative allelic ratios (Pink: maternal; Blue: paternal) are shown under representative chromatograms. H3K27me3 is enriched only on the maternal allele at *Kv-DMR* and on both alleles at the *Inpp5f-v2* and *Plagl1* ICRs, forming thus, with H3K4me2 and H3K4me3 a bivalent chromatin structure on the paternal unmethylated allele. (**B**) Microfluidic-based quantitative RT–PCR analysis of the transcripts initiating from the 15 studied maternal ICRs in BJ1 ES cells. Results are presented as the percentage of the geometrical mean of the expression of the three housekeeping genes *Arbp*, *Gapdh* and *Tbp*. Data were obtained from two independent experiments, each done in duplicate. *Grb10* pat-isoform.

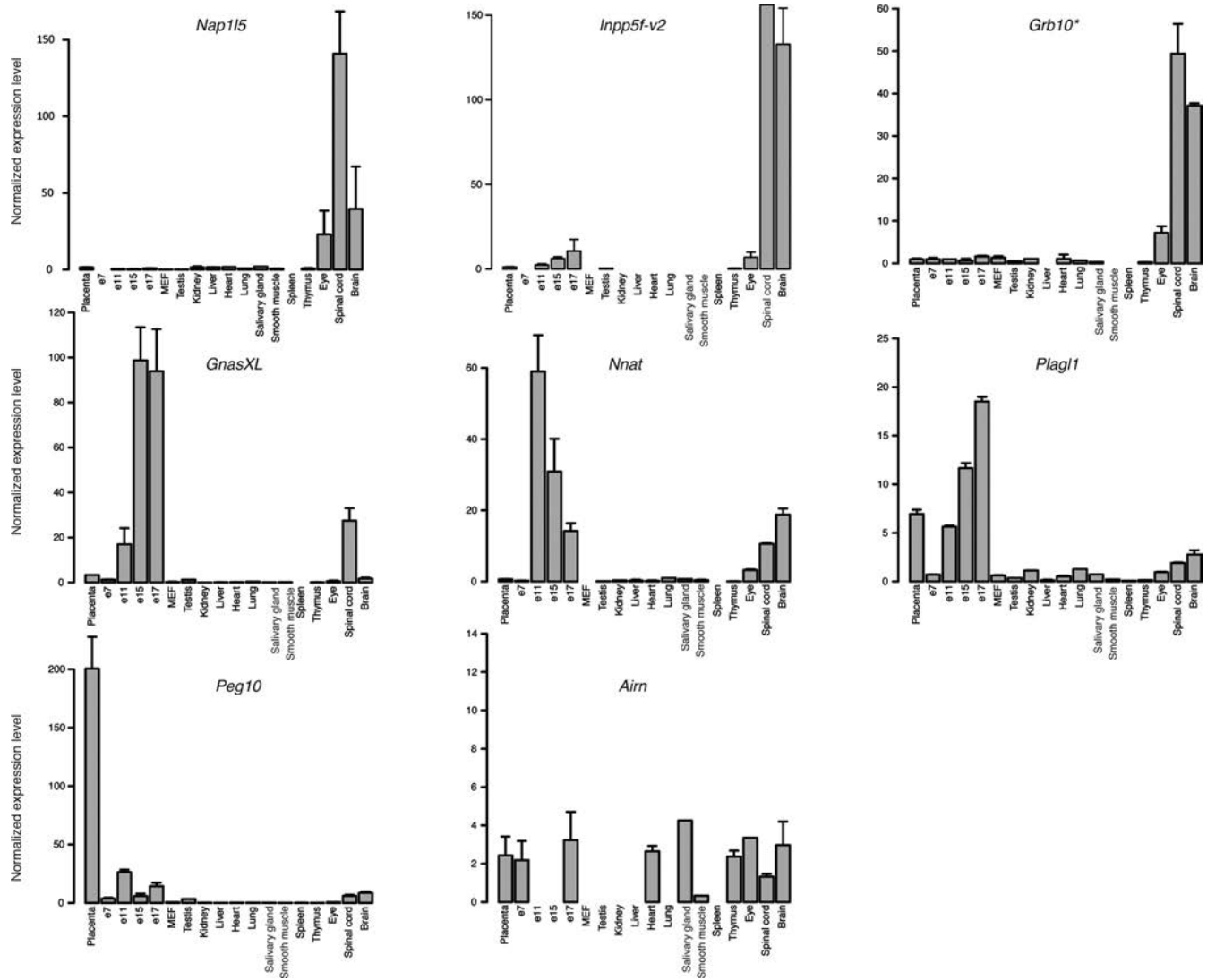

**Figure 2.** Developmental- and tissue-specific gene expression from mat-ICRs associated with bivalent chromatin. Microfluidic-based RT-qPCR analysis of the transcripts initiating from the eight maternal ICRs that are marked by bivalent chromatin in ES cells. Results are presented as the fold enrichment of the mean expression level detected in all tissues, following normalization to the geometrical mean of the expression of the three housekeeping genes *Arbp*, *Gapdh* and *Tbp*. Data were obtained from two independent experiments, each analyzed in duplicate. * *Grb10* pat-isoform.

### Absence of H3K27me3 at the ICR's unmethylated alleles correlates with imprinted tissue-specific expression

The canonical model (25) predicts that, upon differentiation, bivalent domains tend to be resolved either toward H3K4me3 or H3K27me3, leading to stable gene expression activation or repression, respectively. To determine whether this scheme could apply to bivalent chromatin associated with ICRs, we investigated their histone modifications signature in a panel of representative primary cells and tissues (MEFs, E9.5 whole embryos, neonate brain, adult liver and placenta) isolated from reciprocal hybrid C57Bl/6J × JF1 (BJ) and JF1 × C57Bl/6J (JB) mice. In addition to *Grb10*, that we previously analyzed (28), here we conducted detailed analyses at *Nnat*, *Inpp5f-v2*, *Nap1L5* and *Plagl1* loci, which all possess SNPs to assess the parental origin of expression. The expression patterns (assessed by RT-qPCR) recapitulated the results obtained with the microfluidic-based approach. Moreover, we confirmed, using informative SNPs, the imprinted paternal expression in the expressing tissues (Figure 3, Supplementary Figure S4). Following ChIP, we analyzed the allelic distribution of the H3K4me2, H3K4me3, H3K27me3 and H3K9me3 marks. Unexpectedly, at most mat-ICRs, H3K4me3 was maintained irrespectively of the gene expression status. For instance, the *Nnat* and *Inpp5f-v2* DMRs showed a significant paternal allelic enrichment of H3K4me3 in MEFs, liver and placenta where they were not expressed (Figure 3, Supplementary Figure S4B). *Grb10* was an exception as it maintained H3K4me3 only in the expressing tissues (28). We also

Nucleic Acids Research, 2016, Vol. 44, No. 2 627

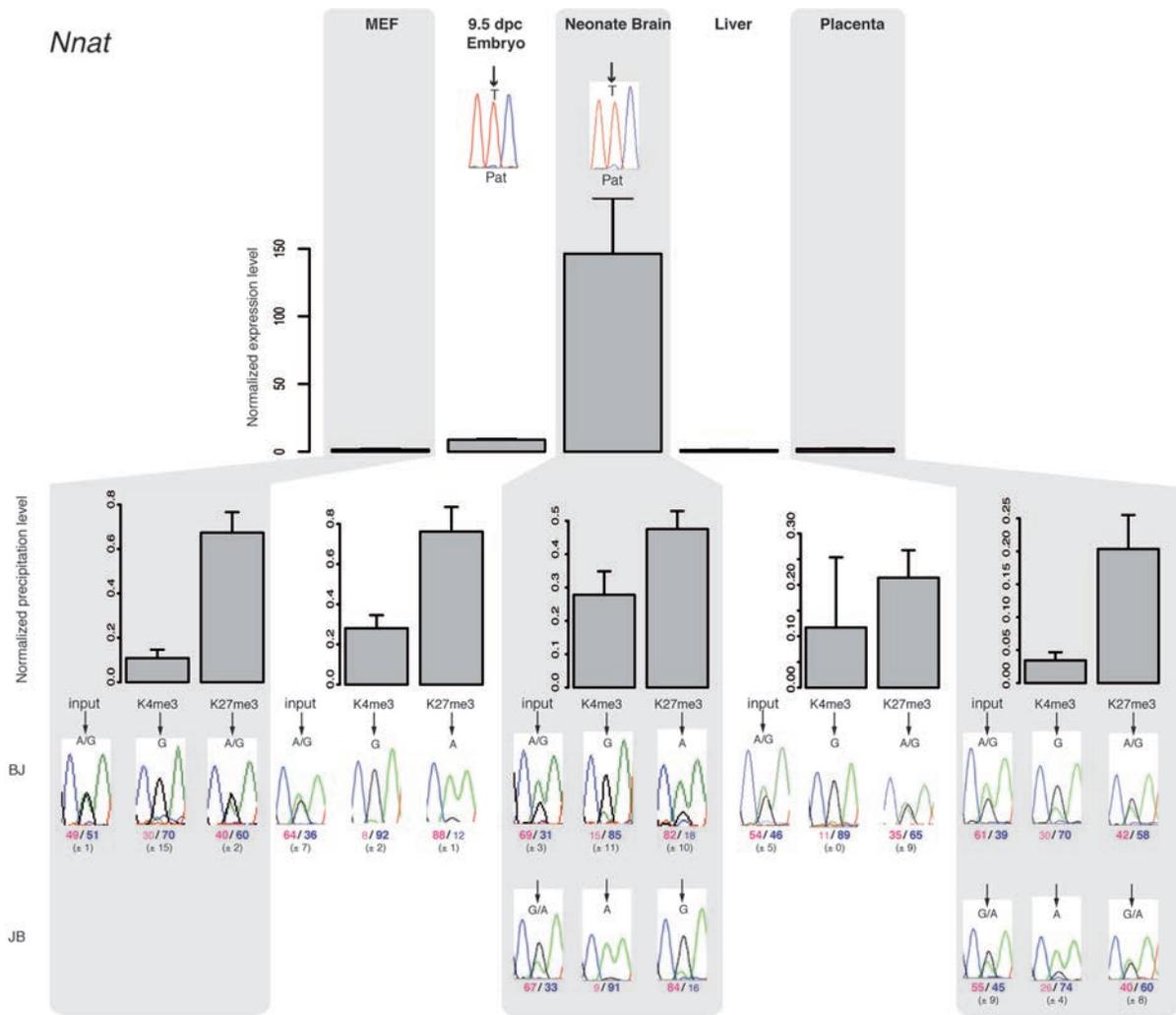

**Figure 3.** Paternal gene expression correlates with loss of H3K27me3 on the paternal allele at bivalent chromatin-associated ICRs. The upper panel shows representative results of gene expression in MEFs, embryos and tissues from BJ mice. Results were normalized to the expression level of the two housekeeping genes *Ppia* and *Rpl30*. Experiments were repeated four times, each in duplicate. The parental origin of gene expression was determined by direct sequencing of the PCR product encompassing a strain-specific SNP in the analyzed region. Lower panels: chromatin analysis following native ChIP using anti-H3-K4me3 and -K27me3 antibodies in BJ and JB (for neonate brain and placenta) samples. The precipitation level was normalized to that obtained at the *Rpl30* promoter (for H3K4me3) and at the *HoxA3* or *HoxD8* promoter (for H3K27me3). The allelic distribution of each mark was determined by direct sequencing of the PCR product encompassing a strain-specific SNP in the analyzed region. The mean values (± standard deviation) of the relative allelic ratios (Pink: maternal; Blue: paternal) are indicated under representative chromatograms. For each tissue, values are the mean of at least three independent ChIP experiments (*n*), each in duplicate: MEFs (*n* = 3); E9.5 embryos (*n* = 3); neonate brain (BJ *n* = 2; JB *n* = 2); liver (*n* = 3); placenta (BJ *n* = 2; JB *n* = 1).

confirmed the previous observation that in all tissue samples and at all tested mat-ICRs, H3K4me2 was associated with the unmethylated paternal allele, while H3K9me3 was consistently associated with the methylated maternal allele (Supplementary Figure S4; 20).

Strikingly, we observed that the H3K27me3 status discriminated between expressing and non-expressing tissues at all analyzed mat-ICRs carrying a bivalent domain in ES cells, but for *Plagl1* (Figure 3, Supplementary Figure S4). For instance, at the *Nnat* ICR, H3K27me3 was absent from the paternal allele only in the two expressing tissues (E9.5 embryos and neonate brain) (Figure 3). Conversely, at the *Plagl1* ICR, H3K27me3, and thus bivalency, was maintained on the paternal allele in neonatal brain and placenta where the gene was paternally expressed (Supplementary Figure S4C). Although we cannot exclude that transcription initiates in an H3K27me3-enriched context or through the use of an alternative promoter, we rather believe that this apparent discrepancy is due to expression arising from a subpopulation of cells within these two tissues. Most of the detected H3K27me3 enrichment would thus be explained by the presence of non-expressing cells. In agreement with this hypothesis, we did not detect H3K27me3 on the paternal allele of the *Plagl1* ICR in MEFs in which *Plagl1* is homogenously expressed. Conversely, in E9.5 embryos, where *Plagl1* expression is restricted to some tissues (52), H3K27me3 was partially enriched on the paternal allele. Moreover, in an *in vitro* system that recapitulates mouse neurogenesis, loss of the H3K27me3 mark from the paternal allele of the *Plagl1* ICR correlated with its de-repression (DN; SMM, PA; in preparation).

Altogether these observations highlight that H3K27me3 removal correlates with tissue-specific expression from the mat-ICRs associated with bivalent chromatin in ES cells, while bivalency is maintained in non-expressing somatic tissues.

### Bivalent chromatin consistently marks mat-ICRs in non-expressing somatic tissues

These results suggest that bivalency could contribute to regulate tissue-specific expression at a subset of mat-ICRs. We thus wanted to know how tissue-specific expression is controlled at the other mat-ICRs that are not associated with bivalent chromatin in ES cells. Microfluidic quantitative RT-PCR analysis using the panel of 17 tissue/embryo samples showed that at some of these ICRs (for instance *KvDMR* and *Zrsr1* (previously known as *U2af1-rs1*), gene expression was widespread (Figure 4 A). Consistently, these ICRs displayed a canonical chromatin signature in all examined tissues, with H3K4me2/me3 only on the paternal allele, as exemplified by the *KvDMR* ICR (Supplementary Figure S5). At other ICRs, paternal tissue-specific expression was observed, although the difference in the expression level between high- and low-expressing tissues was lower compared to what observed at mat-ICRs associated with bivalent chromatin in ES cells (Figure 4A, compare with Figure 2). Unexpectedly, we observed that, again, H3K27me3 enrichment discriminated between expressing (only on the maternal allele) and non-expressing tissues (on both alleles). This indicates that bivalent chromatin is acquired on the paternal allele specifically in non-expressing tissues. For instance, the *Peg3* and *Mest* ICRs both gained H3K27me3 on the paternal allele in MEFs and liver where gene expression was repressed (Figure 4B, Supplementary Figure S6).

Collectively, our results suggest that in non-expressing tissues, bivalent chromatin marks the unmethylated allele of all mat-ICRs, regardless of their chromatin signature in ES cells. This indicates that chromatin bivalency can be gained in somatic tissues through acquisition of H3K27me3.

### Bivalent chromatin is present by default at unmethylated mat-ICRs

To gain insights into the link between bivalency and gene expression at mat-ICRs, we assessed the ICR chromatin signature in $Dnmt3L^{-/+}$ E9.5 embryos obtained from $Dnmt3L^{-/-}$ females, in which DNA methylation imprints at ICRs are not established during oogenesis. In these embryos, lack of maternal DNA methylation at promoter ICRs should lead to biallelic expression of the associated transcript. In agreement with the literature (33,34,53), this was true for most transcripts initiating from mat-ICRs (Figure 5A, Supplementary Figure S7). Conversely, gene expression at the *Grb10* and *Nap1L5* ICRs, which was barely observed in wild type (WT) embryos, was also undetectable in mutant embryos (Figure 5). To investigate the molecular bases of this DNA methylation-independent silencing on the maternal allele, we performed ChIP analyses in WT and mutant embryos. We confirmed and extended our previous observation (20) that in the absence of maternal DNA methylation, allele-specific histone modification signature is lost and both alleles of mat-ICRs adopt a paternal epigenotype. Accordingly, at all analyzed mat-ICRs of $Dnmt3L^{-/+}$ E9.5 embryos, H3K4me2/me3 were enriched on both alleles and the repressive H3K9me3 modification was markedly decreased on the maternal allele (Figure 5). Strikingly, unlike the other repressive marks, H3K27me3 was maintained at the *Nap1l5* and *Grb10* ICRs, while it decreased at the other mat-ICRs (Figure 5B). Combined with the biallelic enrichment for H3K4me2 and H3K4me3, this resulted in the gain of bivalent chromatin also on the maternal allele of these two transcriptionally 'inactive' ICRs.

These results highlight that, unlike the other repressive marks H3K9me3 and H4K20me3 (Figure 5B and (20)), H3K27me3 enrichment on the maternal allele of mat-ICRs is not dependent on DNA methylation, but rather to the absence of transcriptional activity. Collectively, our findings suggest that chromatin bivalency is acquired by default in transcriptionally inactive mat-ICRs.

### Absence of H3K27me3 affects ICR-associated transcription in a locus- and tissue-specific manner

Our observation that bivalency could be the default state of inactive ICRs questions the role of H3K27me3 to dynamically arbitrate expression from these regions, as postulated by the canonical model (25). To determine H3K27me3 functional significance at mat-ICRs, we analyzed the expression of the ICR-associated transcripts in $Eed^{-/-}$ ES cells. EED is a component of the Polycomb repressive complex 2 (PRC2) that mediates H3K27me3 deposition (26). Despite the global reduction of H3K27me3 level (26), gene expression was only mildly affected in $Eed^{-/-}$ ES cells compared to wild type cells (Figure 6A). Specifically, gene expression was reproducibly and significantly increased only at three loci (*Grb10*, *Plagl1* and *Nap1L5*) in $Eed^{-/-}$ compared to WT ES cells, although the expression level at these three 'de-repressed' loci in mutants remained low. Indeed, these transcripts were barely detectable in WT ES cells (Figure 1) and the extent of de-repression (from 2.9- to 5.8 -fold higher) never reached the level observed in somatic tissues.

This observation indicates that in $Eed^{-/-}$ ES cells, loss of H3K27me3 leads to a limited and locus-specific transcription de-repression at bivalent mat-ICRs. This is consistent with a recent study showing that in mouse ES cells deficient in PRC2, only a very small subset of PRC2-target genes are aberrantly expressed (54).

To determine whether this mild effect occurred also in differentiated cells, we analyzed expression in immortalized MEF (iMEF) cells deficient for EZH2, the enzyme of the PRC2 complex that catalyzes H3K27me3. Compared with WT iMEFs, H3K27me3 was markedly decreased at mat-ICRs in $Ezh2^{-/-}$ iMEF cells (Supplementary Figure S8B). However, analysis of the expression of seven ICR-associated transcripts showed that only the expression of three (*Airn*, *Nnat* and *Mest*) was increased in $Ezh2^{-/-}$ iMEFs compared with WT iMEFs, where they were all barely or not detectable (Figure 6B). We next analyzed publicly available RNA-seq data concerning $Ezh2^{-/-}$ car-diomyocytes (37) and spleen regulatory T-cells (Tregs) (38). By focusing our analysis on mat-ICR-associated transcripts that are poorly expressed in WT cells, we observed a very limited and tissue-specific de-repression effect. For instance, *Peg10* transcription was partially de-repressed in $Ezh2^{-/-}$ cardiomyocytes, but not in spleen Tregs (Figure 6C). Similarly, *Nap1L5*, which was deregulated in mutant ES cells, remained unexpressed in $Ezh2^{-/-}$ cardiomyocytes and Tregs (Figure 6C). This is reminiscent of our previous observation that in $Eed^{-/-}$ 6.5 dpc embryos, which lack the PRC2 complex, ectopic paternal expression of *Grb10* was limited to one part of the extra-embryonic ectoderm only, while unaffected in the remaining of the embryo ((28), Supplementary Figure S9).

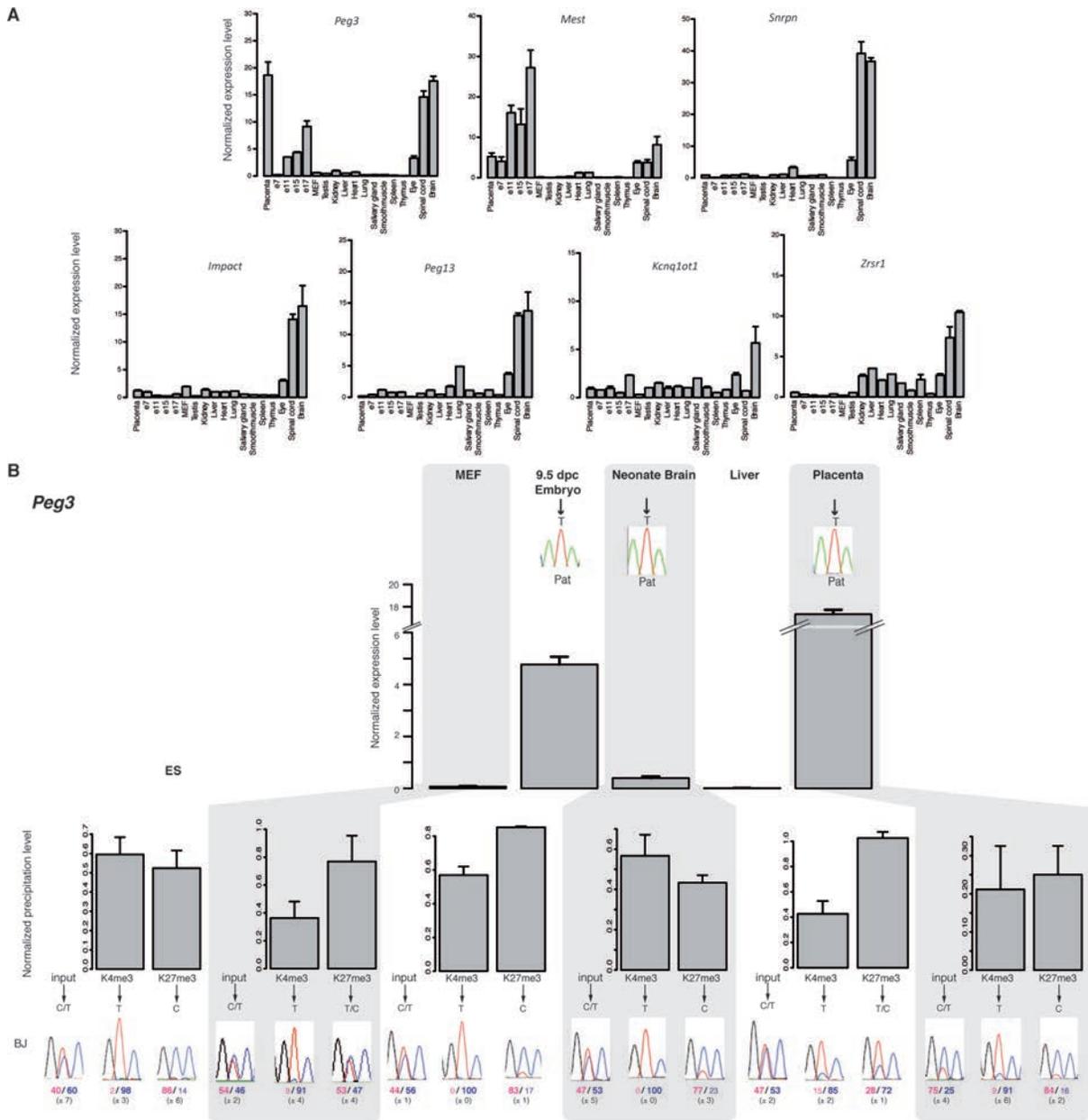

**Figure 4.** Chromatin bivalency marks mat-ICRs in non-expressing tissues. (**A**) Microfluidic-based RT-qPCR analysis of the transcripts initiating from the seven mat-ICRs that are not marked by bivalent chromatin in ES cells. Results are presented as fold enrichment of the mean expression level detected in all tissues, after normalization to the geometrical mean of the expression of the three housekeeping genes *Arbp*, *Gapdh* and *Tbp*. Data were obtained from two independent experiments, each analyzed in duplicate. (**B**) Gene expression and chromatin analysis. Details of the legend are as in Figure 3. The *Peg3* ICR is shown as an example of ICR with tissue-specific promoter activity. The allelic distribution of histone marks was determined by direct sequencing of the PCR product encompassing a strain-specific SNP in the analyzed region. Representative data obtained in BJ samples are shown. The mean values (± standard deviation) of the relative allelic ratios (pink: maternal; Blue: paternal) are given. Chromatin bivalency, which is absent in ES cells, is gained in non-expressing tissues through H3K27me3 acquisition on the paternal allele.

Absence of allelic polymorphisms in $Eed^{-/-}$ ES cells and $Ezh2^{-/-}$ iMEF cells did not allow us to draw formal conclusions about the parental origin of the 'de-repressed' transcripts. Nevertheless, bisulfite analysis showed that DNA methylation at ICRs was similar in WT and mutant cells (Supplementary Figure S8). The detected transcripts were thus likely to originate from the unmethylated paternal allele.

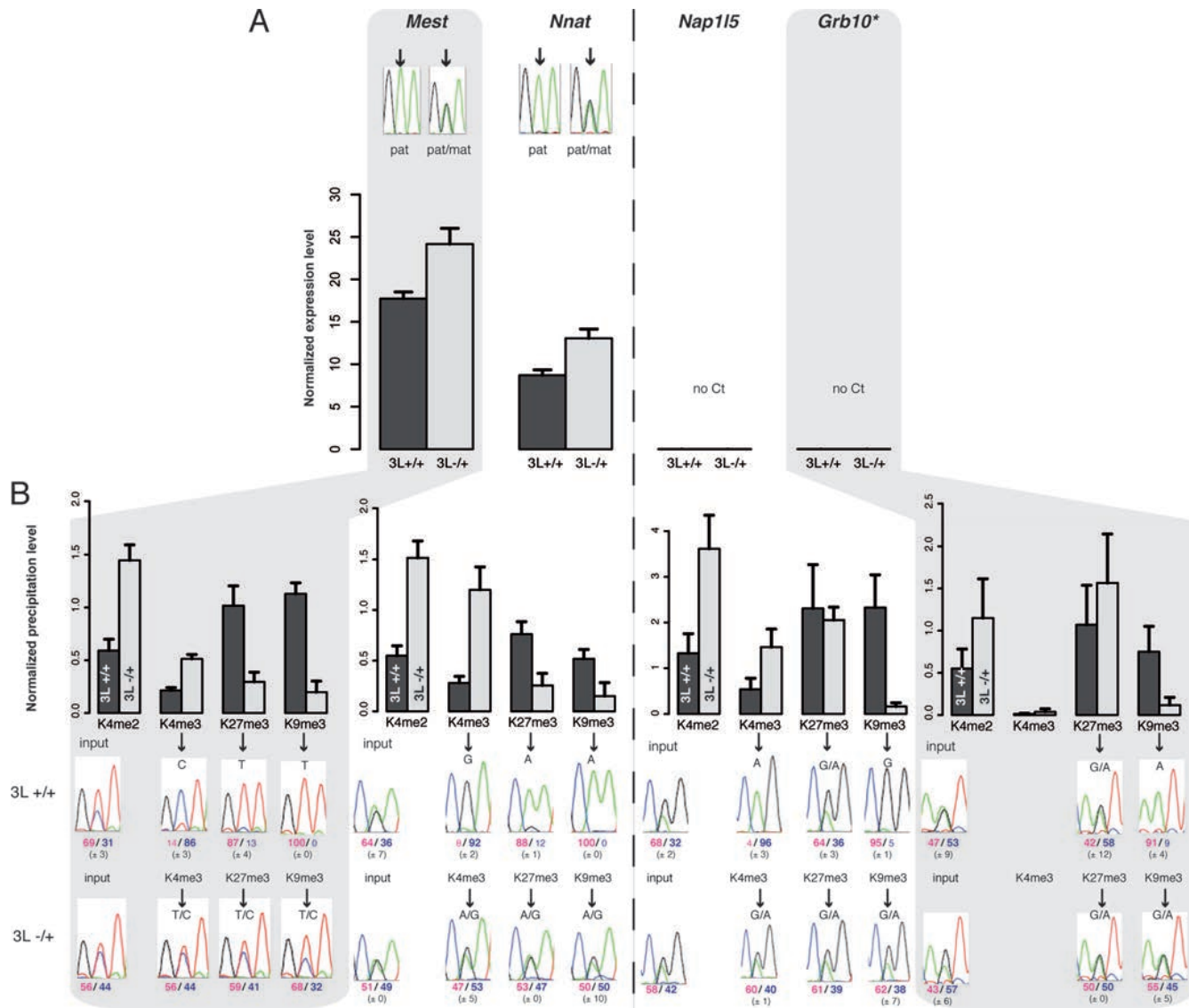

**Figure 5.** Chromatin bivalency is present at both parental alleles of transcriptionally inactive mat-ICRs in *Dnmt3L*$^{-/+}$ embryos. Details of the molecular analyses at ICRs in WT and *Dnmt3l*$^{-/+}$ E9.5 embryos. *Mest* and *Nnat* are representative examples of transcriptionally active and *Nap1L5* and *Grb10* of transcriptionally inactive ICRs in WT embryos. (**A**) Transcriptional analysis using pools of at least 15 WT or mutant embryos. Results are presented as the percentage of expression relative to the geometrical mean of the expression of the two housekeeping genes *Ppia* and *Rpl30*. Data are from two independent experiments, each analyzed in duplicate. No expression was detected from the *Nap1l5* and *Grb10* ICRs (no Ct) in both WT and mutant conceptuses. The parental origin of expression from the *Mest* and *Nnat* ICRs was determined by direct sequencing of the PCR product encompassing a strain-specific SNP in the analyzed region (* *Grb10* pat-isoform). (**B**) Chromatin analysis following native ChIP with anti-H3-K4me2, -K4me3, -K27me3 and -K9me3 antibodies. The precipitation level was normalized to that obtained at the *Rpl30* promoter (for H3K4me2 and H3K4me3), *IAP* (for H3K9me3) and the *Hoxa3* promoter (for H3K27me3). The allelic distribution of these marks was determined by direct sequencing of the PCR product encompassing a strain-specific SNP in the analyzed region. The mean values (± standard deviation) of the relative allelic ratios (pink: maternal; blue: paternal) are indicated under representative chromatograms. Values are the mean of three independent ChIP experiments, each analyzed in duplicate.

Together these findings indicate that absence of H3K27me3 only partially affects the expression of ICR-associated transcripts in a locus- and tissue-specific manner.

## DISCUSSION

The main observation of this study is that H3K27me3 and consequently bivalent chromatin are acquired on the unmethylated allele of mat-ICRs when they are transcriptionally inactive, regardless of the developmental stage or tissue analyzed. Besides suggesting that bivalent chromatin contributes to the fine-tuned regulation of genomic imprinting, this finding provides the proof of concept that the recruitment by default of PRC2 to non-transcribed CpG-island

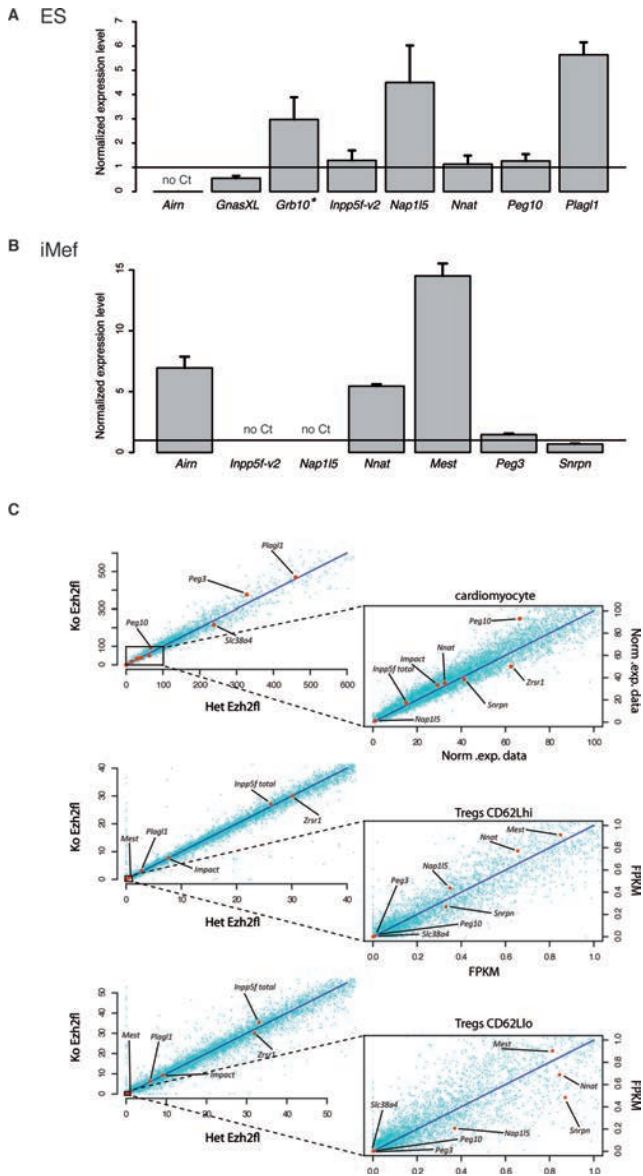

**Figure 6.** Absence of H3K27me3 affects ICR-associated transcript expression in a locus- and tissue-specific manner. (**A** and **B**) Transcription from mat-ICRs associated with bivalent chromatin is mildly affected and only in a subset of loci in $Eed^{-/-}$ ES cells (A) and $Ezh2^{-/-}$ iMEF cells (B). The results of RT-qPCR analyses are presented as the fold enrichment of the expression level obtained in $Eed^{+/+}$ ES and $Ezh2^{+/+}$ iMEF cells, respectively. Data are from four independent experiments, each analyzed in duplicate. (**C**) Transcription from maternal ICRs associated with bivalent chromatin in $Ezh2^{-/-}$ somatic cells. Publicly available RNA-seq data were used to produce scatter plots to correlate genome-wide the normalized expression between heterozygotes controls (Het Ezh2 fl) and Ezh2 mutant (Ko Ezh2 fl) E12.5 cardiomyocytes, resting (CD62Lhi) and activated (CD62Llo) spleen T regulatory cells (Tregs). Transcripts initiating from the ICRs are shown by red dots. *Peg10* expression is partly affected in $Ezh2^{-/-}$ cardiomyocytes, but not in $Ezh2^{-/-}$ Tregs cells compared to controls ($Ezh2^{+/-}$ cells).

genes, so far described only in pluripotent cells (54,55), can also occur in somatic lineages.

Initially, it was hypothesized that bivalent chromatin was a pluripotency-associated feature to poise genes for activation or repression upon differentiation (25). Consistently, widespread resolution of bivalency has been observed upon ES cell differentiation, although some loci retain bivalency, indicating that non-pluripotent cells can also contain bivalent domains (16,56). The genome-wide dynamics of chromatin bivalency upon differentiation was studied in an *in vitro* model of neurogenesis in which stem cells differentiate into neuronal progenitors and finally into glutamatergic pyramidal neurons (17). Strikingly, the authors observed that several hundred '*de novo*' bivalent domains are gained in neural progenitors and again upon terminal differentiation. Our findings in mat-ICRs also support the hypothesis that chromatin bivalency is not limited to ES cells and further confirm that this chromatin signature can be acquired at all development stages. This is well-illustrated, for instance, by the *Mest* and *Peg3* ICRs that are transcriptionally active in ES cells and that acquire bivalency in non-expressing somatic tissues.

However, at mat-ICRs this gain relies on acquisition of the H3K27me3 mark only. Indeed, we observed that the paternal unmethylated allele of ICRs is marked by H3K4me3 at most developmental stages, regardless of their transcriptional activity, and further confirmed the constitutive enrichment for H3K4me2 (20). Consequently, we did not find an H3K27me3-only signature, as observed at other bivalent genes, following resolution toward transcription repression (16). Genome-wide studies suggest that this feature is probably not limited to ICRs, but rather linked to the CpG-richness of the region. Most CpG island promoters indeed show enrichment for H3K4me2 and H3K4me3, regardless of the associated transcriptional activity and the cell differentiation state (16,17,57).

Recent studies have shown that in ES cells, H3K4me3 at bivalent promoters is controlled by MLL2, one of six SET1/trithorax-type H3K4 methyltransferases (58,59) that are thought to be recruited to unmethylated CpG islands without assistance of transcription factors (59). One might expect that MLL2, which is widely expressed during development and in adult tissues (60), could also account for the maintenance of the H3K4me2/3 marks at bivalent domain-associated ICRs in somatic tissues. The mechanism involved in establishing the PRC2-mediated H3K27me3 remains more elusive because this complex does not contain any DNA binding domain (27). In the traditional 'instructive model', supported by several studies, transcription factors or long non-coding RNAs are instrumental in promoting PCR2 recruitment to chromatin (e.g. 61,62). However, this model hardly accounts for the observation that bivalency is the default chromatin signature at artificially introduced CpG islands in ES cells (63–65). A second model ('responsive model') predicts that PRC2 recruitment is responsive to permissive chromatin signatures at CpG islands (55). A recent elegant study supports this hypothesis by showing that, in ES cells, PRC2 is recruited by default at transcriptionally inactive CpG islands (54). The recent discovery that the hierarchical model whereby PRC1 is recruited in a PRC2-dependent manner might be reversed further provides a mechanism to explain the establishment of H3K27me3 by default. In the emerging model, a CxxC domain-containing factor that specifically recognizes unmethylated DNA, such as KDM2B, recruits PRC1 to unmethylated and transcriptionally inactive CpG islands, leading to H2AK119ub1 deposition. Subsequently this modification will recruit PRC2, by a yet to be established mechanism, to ultimately catalyze H3K27me3 deposition at CpG islands (66–69).

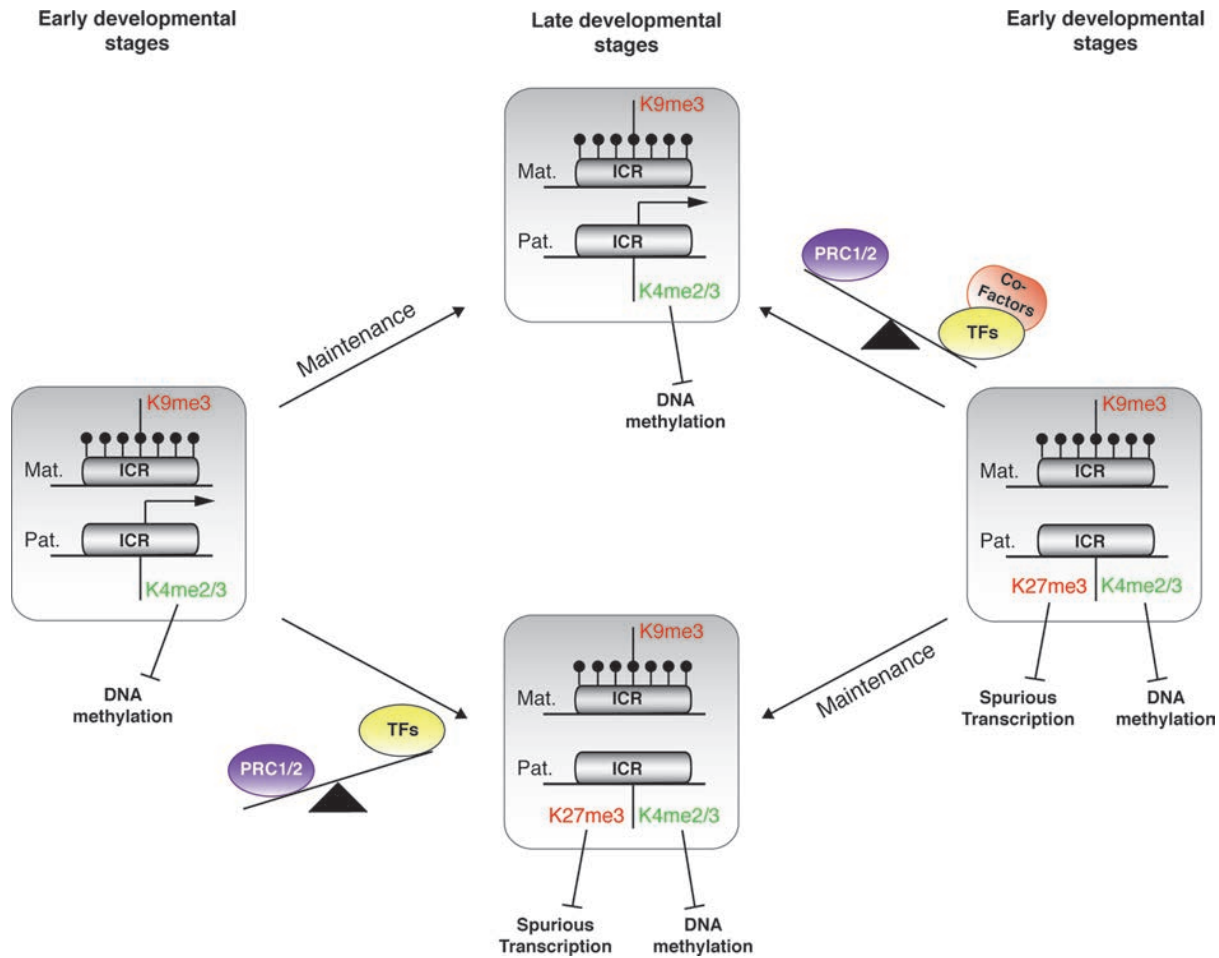

**Figure 7.** Model of the fine-tuned regulation of imprinted gene expression by chromatin bivalency. Model illustrating the gain and loss of chromatin bivalency at maternal ICRs during development. Relative to their transcriptional activity, the ICR paternal allele can be associated (right panel) or not (left panel) with bivalent chromatin in pluripotent cells (early developmental stages). This signature will be maintained, upon differentiation, in lineage sharing the same transcriptional status (late developmental stages). During lineage commitment, in the absence of transcriptional activity or transcription below a certain threshold, H3K27me3 is deposited by default on the paternal allele, already marked by H3K4me2/3, thus leading to the formation of a bivalent domain (from left to lower central panel). This signature, by repelling DNA methylation and protecting against spurious expression, contributes to protecting the ICR integrity in non-expressing cells. Conversely, the switch from gene repression to expression, upon lineage commitment, might rely on the availability of the *ad hoc* transcription machinery associated with H3K27me3 removal. Specific co-factors, for instance H3K27me3 demethylases, could counteract PRC1/2 action and ensure robust transcription from the paternal allele (from right to upper central panel). DNA methylation is represented by filled lollipops. Arrows indicate active transcription.

Our study provides the proof of concept that the 'responsive model', so far supported by experiments in ES cells and, in part, using artificially introduced CpG islands, can also apply *in vivo* and during the key developmental process that is genomic imprinting. A broader characterization of the epigenetic signature of ICRs in a panel of expressing and non-expressing tissues, relative to H2AK119ub1 presence for instance, could provide clues about the underlying mechanism.

The deposition by default of H3K27me3 at ICRs in non-expressing tissues challenges the expected role of bivalency to arbitrate gene expression upon differentiation and development. In PRC2 deficient cells and tissues, we observed that gene transcription was increased only in a subset of mat-ICRs and only slightly, further supporting the hypothesis that H3K27me3 removal on its own is not sufficient to trigger sustained transcription de-repression at ICRs. Also, removal of H3K27me3 in ES cells has a limited transcriptional impact genome-wide (54). Similarly, loss of H3K4me3 at bivalent domains in ES cells lacking MLL2 has a very mild effect on the expression of the associated genes upon differentiation (58,59).

Altogether, these findings indicate that the function of chromatin bivalency is much less dynamic than expected based on the canonical model. Our observations rather suggest that the primary role of bivalency is rather protective at maternally methylated ICRs to ensure that the paternal allele remains unmethylated, although repressed in a subset of tissues. Specifically, in our working hypothesis, absence of transcriptional activity or transcription below a certain threshold is sufficient to allow PRC1/2 recruitment to the paternal unmethylated allele of ICRs and the subsequent deposition of H3K27me3 in somatic lineages (Figure 7). By contrast H3K4me2 and possibly also H3K4me3 are maintained during the entire development in all somatic lineages through MLL2, when present in a bivalent configuration, and by another transcription factor-associated MLL/SET1 complex at transcriptionally active ICRs. Importantly, as the DNA methylation complex is repelled by H3K4me2/3 (9,70), this configuration ensures that the paternal allele of mat-ICRs remains unmethylated during development. Concomitantly, H3K27me3 protects against spurious and unscheduled transcription that could initiate in non-expressing tissues that partially share the pool of transcription factors found in expressing tissues but unable to sustain an appropriate expression level (Figure 7). This model highlights that two silencing mechanisms are functional at ICRs in an allele-specific fashion. The first one, on the maternal allele, is constitutive and relies on DNA methylation and repressive histone marks. The second one, based on bivalency, contributes to the appropriate tissue-specific expression from the paternal allele.

According to this model, the switch from gene repression to expression during lineage commitment would rely on transcription-induced loss of H3K27me3 at ICRs, as suggested by the loss of H3K27me3 at the expressed allele, but not from the silent maternally methylated allele. Although the underlying mechanism remains elusive, one might hypothesize that in the lineage possessing the *ad hoc* transcription factors, the assembled transcription machinery might trigger the loss of H3K27me3, by recruiting histone demethylases, for instance, and then ensure a robust transcription. Further analyses at ICRs in a controlled system of cell/tissue differentiation should provide insights on this mechanism and more broadly on the control of bivalent domain resolution during lineage commitment.

Finally, our observation is also relevant for understanding the molecular bases of imprinting disorders. Genetic alterations (such as uniparental disomy, chromosome imbalance and point mutations) or abnormal DNA methylation at ICRs are causally linked to most imprinting disorder cases; however, some phenotypically well-characterized patients do no display any of these molecular defects. This is observed for instance in ∼10% of patients with Angelman syndrome, a severe neurobehavioral disorder, and in 40% of patients with Silver-Russell syndrome, a pre- and postnatal growth disorder. Alteration in the control of bivalency at ICRs could help understanding these cases. Indeed, the tissue- and locus-specific effects we observed in PRC2-deficient samples suggests that aberrant loss of H3K27me3 at ICRs could contribute to the somatic mosaicism and to the heterogeneous clinical features widely documented in these pathologies (71).

Our study shows that chromatin bivalency contributes to the fine-tuned regulation of imprinted gene expression. Specifically, we propose that it plays primarily a safe-guard function to ensure that the paternal allele of maternal ICRs remains constitutively unmethylated, while being firmly silent in a subset of tissues. In addition, its developmentally controlled resolution, through the loss of H3K27me3, ensures robust gene expression in the appropriate cell types and developmental stages. Beyond imprinting, our study provides the proof of concept that bivalency can be the default state of transcriptionally inactive CpG islands/promoters, regardless of the developmental stage. This raises the question of whether the protective model we hypothesized at ICRs could be extended genome-wide to protect cell identity.

## ACCESSION NUMBER

DRA accession number: DRX005582 (http://trace.ddbj.nig.ac.jp/dra/index_e.html).

## SUPPLEMENTARY DATA

Supplementary Data are available at NAR Online.

## ACKNOWLEDGEMENTS

We thank A. Fisher, H. Jorgensen and V. Azuara for the generous gift of DNA and RNA from *Eed*$^{(-/-)}$ ES cell lines, Kohji Okamura for his assistance to extract JF1 SNPs and the Platform "Gentyane" (http://gentyane.clermont.inra.fr/) for its technical help with the Fluidigm 96.96 Dynamic Arrays . We also thank Drs David Monk, Isabelle Vaillant and Tristan Bouschet and all members of PA's team for critical reading of the manuscript.

## FUNDING

Agence National de la Recherche [ANR 2010 JCJC 1601 'Bivandev']; Plan Cancer-INSERM [C14085CS 'Gliobiv']; La Ligue Contre le Cancer comité Puy de Dôme et Ardéche; 'Conseil regional d'Auvergne' (to P.A.); Fondation ARC pour la Recherche sur le Cancer fellowship [PDF20120604971] (to S.M.M.); Université d'Auvergne fellowship (to F.C.). Funding for open access charge: CNRS (Centre National de la Recherche Scientifique).
*Conflict of interest statement.* None declared.

## REFERENCES

1. Peters,J. (2014) The role of genomic imprinting in biology and disease: an expanding view. *Nat. Rev. Genet.*, **15**, 517–530.
2. Plasschaert,R.N. and Bartolomei,M.S. (2014) Genomic imprinting in development, growth, behavior and stem cells. *Development*, **141**, 1805–1813.
3. Arnaud,P. and Feil,R. (2005) Epigenetic deregulation of genomic imprinting in human disorders and following assisted reproduction. *Birth Defects Res. C Embryo Today*, **75**, 81–97.
4. Arnaud,P. (2010) Genomic imprinting in germ cells: imprints are under control. *Reproduction*, **140**, 411–423.
5. Adalsteinsson,B.T. and Ferguson-Smith,A.C. (2014) Epigenetic control of the genome-lessons from genomic imprinting. *Genes*, **5**, 635–655.
6. Kelsey,G. and Feil,R. (2013) New insights into establishment and maintenance of DNA methylation imprints in mammals. *Philos. Trans. R Soc. Lond. B Biol. Sci.*, **368**, 20110336.
7. Ciccone,D.N., Su,H., Hevi,S., Gay,F., Lei,H., Bajko,J., Xu,G., Li,E. and Chen,T. (2009) KDM1B is a histone H3K4 demethylase required to establish maternal genomic imprints. *Nature*, **461**, 415–418.
8. Chotalia,M., Smallwood,S.A., Ruf,N., Dawson,C., Lucifero,D., Frontera,M., James,K., Dean,W. and Kelsey,G. (2009) Transcription is required for establishment of germline methylation marks at imprinted genes. *Genes Dev.*, **23**, 105–117.

9. Ooi,S.K., Qiu,C., Bernstein,E., Li,K., Jia,D., Yang,Z., Erdjument-Bromage,H., Tempst,P., Lin,S.P., Allis,C.D. *et al.* (2007) DNMT3L connects unmethylated lysine 4 of histone H3 to de novo methylation of DNA. *Nature*, **448**, 714–717.
10. Li,X., Ito,M., Zhou,F., Youngson,N., Zuo,X., Leder,P. and Ferguson-Smith,A.C. (2008) A maternal-zygotic effect gene, Zfp57, maintains both maternal and paternal imprints. *Dev. Cell*, **15**, 547–557.
11. Nakamura,T., Arai,Y., Umehara,H., Masuhara,M., Kimura,T., Taniguchi,H., Sekimoto,T., Ikawa,M., Yoneda,Y., Okabe,M. *et al.* (2007) PGC7/Stella protects against DNA demethylation in early embryogenesis. *Nat. Cell Biol.*, **9**, 64–71.
12. Quenneville,S., Verde,G., Corsinotti,A., Kapopoulou,A., Jakobsson,J., Offner,S., Baglivo,I., Pedone,P.V., Grimaldi,G., Riccio,A. *et al.* (2011) In embryonic stem cells, ZFP57/KAP1 recognize a methylated hexanucleotide to affect chromatin and DNA methylation of imprinting control regions. *Mol. Cell*, **44**, 361–723.
13. Li,E., Beard,C. and Jaenisch,R. (1993) Role for DNA methylation in genomic imprinting. *Nature*, **366**, 362–365.
14. Hirasawa,R., Chiba,H., Kaneda,M., Tajima,S., Li,E., Jaenisch,R. and Sasaki,H. (2008) Maternal and zygotic Dnmt1 are necessary and sufficient for the maintenance of DNA methylation imprints during preimplantation development. *Genes Dev.*, **22**, 1607–1616.
15. Barski,A., Cuddapah,S., Cui,K., Roh,T.Y., Schones,D.E., Wang,Z., Wei,G., Chepelev,I. and Zhao,K. (2007) High-resolution profiling of histone methylations in the human genome. *Cell*, **129**, 823–837.
16. Mikkelsen,T.S., Ku,M., Jaffe,D.B., Issac,B., Lieberman,E., Giannoukos,G., Alvarez,P., Brockman,W., Kim,T.K., Koche,R.P. *et al.* (2007) Genome-wide maps of chromatin state in pluripotent and lineage-committed cells. *Nature*, **448**, 553–560.
17. Mohn,F., Weber,M., Rebhan,M., Roloff,T.C., Richter,J., Stadler,M.B., Bibel,M. and Schübeler,D. (2008) Lineage-specific polycomb targets and de novo DNA methylation define restriction and potential of neuronal progenitors. *Mol. Cell*, **30**, 755–766.
18. Zhang,Y., Jurkowska,R., Soeroes,S., Rajavelu,A., Dhayalan,A., Bock,I., Rathert,P., Brandt,O., Reinhardt,R., Fischle,W. *et al.* (2010) Chromatin methylation activity of Dnmt3a and Dnmt3a/3L is guided by interaction of the ADD domain with the histone H3 tail. *Nucleic Acids Res.*, **38**, 4246–4253.
19. Dhayalan,A., Rajavelu,A., Rathert,P., Tamas,R., Jurkowska,R.Z., Ragozin,S. and Jeltsch,A. (2010) The Dnmt3a PWWP domain reads histone 3 lysine 36 trimethylation and guides DNA methylation. *J. Biol. Chem.*, **285**, 26114–26120.
20. Henckel,A., Nakabayashi,K., Sanz,L.A., Feil,R., Hata,K. and Arnaud,P. (2009) Histone methylation is mechanistically linked to DNA methylation at imprinting control regions in mammals. *Hum. Mol. Genet.*, **18**, 3375–3383.
21. Henckel,A., Chebli,K., Kota,S.K., Arnaud,P. and Feil,R. (2012) Transcription and histone methylation changes correlate with imprint acquisition in male germ cells. *EMBO J.*, **31**, 606–615.
22. Singh,P., Li,A.X., Tran,D.A., Oates,N., Kang,E.R., Wu,X. and Szabó,P.E. (2013) De novo DNA methylation in the male germ line occurs by default but is excluded at sites of H3K4 methylation. *Cell Rep.*, **4**, 205–219.
23. Ginno,P.A., Lott,P.L., Christensen,H.C., Korf,I. and Chédin,F. (2012) R-loop formation is a distinctive characteristic of unmethylated human CpG island promoters. *Mol. Cell*, **45**, 814–825.
24. Krajewski,W.A., Nakamura,T., Mazo,A. and Canaani,E. (2005) A motif within SET-domain proteins binds single-stranded nucleic acids and transcribed and supercoiled DNAs and can interfere with assembly of nucleosomes. *Mol. Cell. Biol.*, **25**, 1891–1899.
25. Bernstein,B.E., Mikkelsen,T.S., Xie,X., Kamal,M., Huebert,D.J., Cuff,J., Fry,B., Meissner,A., Wernig,M., Plath,K. *et al.* (2006) A bivalent chromatin structure marks key developmental genes in embryonic stem cells. *Cell*, **125**, 315–326.
26. Azuara,V., Perry,P., Sauer,S., Spivakov,M., Jørgensen,H.F., John,R.M., Gouti,M., Casanova,M., Warnes,G., Merkenschlager,M. *et al.* (2006) Chromatin signatures of pluripotent cell lines. *Nat. Cell Biol.*, **8**, 532–538.
27. Voigt,P., Tee,W.W. and Reinberg,D. (2013) A double take on bivalent promoters. *Genes Dev.*, **27**, 1318–1338.
28. Sanz,L.A., Chamberlain,S., Sabourin,J.C., Henckel,A., Magnuson,T., Hugnot,J.P., Feil,R. and Arnaud,P. (2008) A mono-allelic bivalent chromatin domain controls tissue-specific imprinting at Grb10. *EMBO J.*, **27**, 2523–2532.
29. Monk,D., Arnaud,P., Frost,J., Hills,F.A., Stanier,P., Feil,R. and Moore,G.E. (2009) Reciprocal imprinting of human GRB10 in placental trophoblast and brain: evolutionary conservation of reversed allelic expression. *Hum. Mol. Genet.*, **18**, 3066–3074.
30. Verona,R.I., Thorvaldsen,J.L., Reese,K.J. and Bartolomei,M.S. (2008) The transcriptional status but not the imprinting control region determines allele-specific histone modifications at the imprinted H19 locus. *Mol. Cell. Biol.*, **28**, 71–82.
31. Han,L., Lee,D.H. and Szabó,P.E. (2008) CTCF is the master organizer of domain-wide allele-specific chromatin at the H19/Igf2 imprinted region. *Mol. Cell. Biol.*, **28**, 1124–1135.
32. McEwen,K.R. and Ferguson-Smith,A.C. (2010) Distinguishing epigenetic marks of developmental and imprinting regulation. *Epigenet. Chromatin*, **3**, 2.
33. Hata,K., Okano,M., Lei,H. and Li,E. (2002) Dnmt3L cooperates with the Dnmt3 family of de novo DNA methyltransferases to establish maternal imprints in mice. *Development*, **129**, 1983–1993.
34. Arnaud,P., Hata,K., Kaneda,M., Li,E., Sasaki,H., Feil,R. and Kelsey,G. (2006) Stochastic imprinting in the progeny of Dnmt3L−/− females. *Hum. Mol. Genet.*, **15**, 589–598.
35. Livak,K.J. and Schmittgen,T.D. (2001) Analysis of relative gene expression data using real-time quantitative PCR and the 2(-Delta Delta C(T)) Method. *Methods*, **25**, 402–408.
36. Wagschal,A., Delaval,K., Pannetier,M., Arnaud,P. and Feil,R. (2007) PCR-based analysis of immunoprecipitated chromatin. *CSH Protoc.*, 2007:pdb.prot4768.
37. He,A., Ma,Q., Cao,J., von Gise,A., Zhou,P., Xie,H., Zhang,B., Hsing,M., Christodoulou,D.C., Cahan,P. *et al.* (2012) Polycomb repressive complex 2 regulates normal development of the mouse heart. *Circ. Res.*, **110**, 406–415.
38. DuPage,M., Chopra,G., Quiros,J., Rosenthal,W.L., Morar,M.M., Holohan,D., Zhang,R., Turka,L., Marson,A. and Bluestone,J.A. (2015) The chromatin-modifying enzyme ezh2 is critical for the maintenance of regulatory T cell identity after activation. *Immunity*, **42**, 227–238.
39. Kota,S.K., Llères,D., Bouschet,T., Hirasawa,R., Marchand,A., Begon-Pescia,C., Sanli,I., Arnaud,P., Journot,L., Girardot,M. *et al.* (2014) ICR noncoding RNA expression controls imprinting and DNA replication at the Dlk1-Dio3 domain. *Dev. Cell*, **31**, 19–33.
40. Henckel,A. and Arnaud,P. (2010) Genome-wide identification of new imprinted genes. *Brief. Funct. Genomics*, **9**, 304–314.
41. Brinkman,A.B., Gu,H., Bartels,S.J., Zhang,Y., Matarese,F., Simmer,F., Marks,H., Bock,C., Gnirke,A., Meissner,A. *et al.* (2012) Sequential ChIP-bisulfite sequencing enables direct genome-scale investigation of chromatin and DNA methylation cross-talk. *Genome Res.*, **22**, 1128–1138.
42. Statham,A.L., Robinson,M.D., Song,J.Z., Coolen,M.W., Stirzaker,C. and Clark,S.J. (2012) Bisulfite sequencing of chromatin immunoprecipitated DNA (BisChIP-seq) directly informs methylation status of histone-modified DNA. *Genome Res.*, **22**, 1120–1127.
43. Umlauf,D., Goto,Y., Cao,R., Cerqueira,F., Wagschal,A., Zhang,Y. and Feil,R. (2004) Imprinting along the Kcnq1 domain on mouse chromosome 7 involves repressive histone methylation and recruitment of Polycomb group complexes. *Nat. Genet.*, **36**, 1296–1300.
44. Lewis,A., Green,K., Dawson,C., Redrup,L., Huynh,K.D., Lee,J.T., Hemberger,M. and Reik,W. (2006) Epigenetic dynamics of the Kcnq1 imprinted domain in the early embryo. *Development*, **133**, 4203–4210.
45. Monk,D., Arnaud,P., Apostolidou,S., Hills,F.A., Kelsey,G., Stanier,P., Feil,R. and Moore,G.E. (2006) Limited evolutionary conservation of imprinting in the human placenta. *Proc. Natl. Acad. Sci. U.S.A.*, **103**, 6623–6628.
46. Terranova,R., Yokobayashi,S., Stadler,M.B., Otte,A.P., van Lohuizen,M., Orkin,S.H. and Peters,A.H. (2008) Polycomb group proteins Ezh2 and Rnf2 direct genomic contraction and imprinted repression in early mouse embryos. *Dev. Cell*, **15**, 668–679.
47. Kim,J.M. and Ogura,A. (2009) Changes in allele-specific association of histone modifications at the imprinting control regions during mouse preimplantation development. *Genesis*, **47**, 611–616.
48. Di Cerbo,V., Mohn,F., Ryan,D.P., Montellier,E., Kacem,S., Tropberger,P., Kallis,E., Holzner,M., Hoerner,L., Feldmann,A. *et al.*

48. (2014) Acetylation of histone H3 at lysine 64 regulates nucleosome dynamics and facilitates transcription. *Elife*, **25**, e01632.
49. Brookes,E., de Santiago,I., Hebenstreit,D., Morris,K.J., Carroll,T., Xie,S.Q., Stock,J.K., Heidemann,M., Eick,D., Nozaki,N. *et al.* (2012) Polycomb associates genome wide with a specific RNA polymerase II variant, and regulates metabolic genes in ESCs. *Cell Stem Cell*, **10**, 157–170.
50. Smith,R.J., Dean,W., Konfortova,G. and Kelsey,G. (2003) Identification of novel imprinted genes in a genome-wide screen for maternal methylation. *Genome Res.*, **13**, 558–569.
51. Choi,J.D., Underkoffler,L.A., Wood,A.J., Collins,J.N., Williams,P.T., Golden,J.A., Schuster,E.F. Jr, Loomes,K.M. and Oakey,R.J. (2005) A novel variant of Inpp5f is imprinted in brain, and its expression is correlated with differential methylation of an internal CpG island. *Mol. Cell. Biol.*, **25**, 5514–5522.
52. Valente,T. and Auladell,C. (2001) Expression pattern of Zac1 mouse gene, a new zinc-finger protein that regulates apoptosis and cellular cycle arrest, in both adult brain and along development. *Mech. Dev.*, **108**, 207–211.
53. Bourc'his,D., Xu,G.L., Lin,C.S., Bollman,B. and Bestor,T.H. (2001) Dnmt3L and the establishment of maternal genomic imprints. *Science*, **294**, 2536–2539.
54. Riising,E.M., Comet,I., Leblanc,B., Wu,X., Johansen,J.V. and Helin,K. (2014) Gene silencing triggers polycomb repressive complex 2 recruitment to CpG islands genome wide. *Mol. Cell*, **55**, 347–360.
55. Klose,R.J., Cooper,S., Farcas,A.M., Blackledge,N.P. and Brockdorff,N. (2013) Chromatin sampling–an emerging perspective on targeting polycomb repressor proteins. *PLoS Genet.*, **9**, e1003717.
56. Pan,G., Tian,S., Nie,J., Yang,C., Ruotti,V., Wei,H., Jonsdottir,G.A., Stewart,R. and Thomson,J.A. (2007) Whole-genome analysis of histone H3 lysine 4 and lysine 27 methylation in human embryonic stem cells. *Cell Stem Cell*, **1**, 299–312.
57. Thomson,J.P., Skene,P.J., Selfridge,J., Clouaire,T., Guy,J., Webb,S., Kerr,A.R., Deaton,A., Andrews,R., James,K.D. *et al.* (2010) CpG islands influence chromatin structure via the CpG-binding protein Cfp1. *Nature*, **464**, 1082–1086.
58. Hu,D., Garruss,A.S., Gao,X., Morgan,M.A., Cook,M., Smith,E.R. and Shilatifard,A. (2013) The Mll2 branch of the COMPASS family regulates bivalent promoters in mouse embryonic stem cells. *Nat. Struct. Mol. Biol.*, **20**, 1093–1097.
59. Denissov,S., Hofemeister,H., Marks,H., Kranz,A., Ciotta,G., Singh,S., Anastassiadis,K., Stunnenberg,H.G. and Stewart,A.F. (2014) Mll2 is required for H3K4 trimethylation on bivalent promoters in embryonic stem cells, whereas Mll1 is redundant. *Development*, **141**, 526–537.
60. Gu,B. and Lee,M.G. (2013) Histone H3 lysine 4 methyltransferases and demethylases in self-renewal and differentiation of stem cells. *Cell Biosci.*, **3**, 39.
61. Rinn,J.L., Kertesz,M., Wang,J.K., Squazzo,S.L., Xu,X., Brugmann,S.A., Goodnough,L.H., Helms,J.A., Farnham,P.J., Segal,E. *et al.* (2007) Functional demarcation of active and silent chromatin domains in human HOX loci by noncoding RNAs. *Cell*, **129**, 1311–1323.
62. Palacios,D., Mozzetta,C., Consalvi,S., Caretti,G., Saccone,V., Proserpio,V., Marquez,V.E., Valente,S., Mai,A., Forcales,S.V. *et al.* (2010) TNF/p38α/polycomb signaling to Pax7 locus in satellite cells links inflammation to the epigenetic control of muscle regeneration. *Cell Stem Cell*, **7**, 455–469.
63. Mendenhall,E.M., Koche,R.P., Truong,T., Zhou,V.W., Issac,B., Chi,A.S., Ku,M. and Bernstein,B.E. (2010) GC-rich sequence elements recruit PRC2 in mammalian ES cells. *PLoS Genet.*, **6**, e1001244.
64. Lynch,M.D., Smith,A.J., De Gobbi,M., Flenley,M., Hughes,J.R., Vernimmen,D., Ayyub,H., Sharpe,J.A., Sloane-Stanley,J.A., Sutherland,L. *et al.* (2012) An interspecies analysis reveals a key role for unmethylated CpG dinucleotides in vertebrate Polycomb complex recruitment. *EMBO J.*, **31**, 317–329.
65. Wachter,E., Quante,T., Merusi,C., Arczewska,A., Stewart,F., Webb,S. and Bird,A. (2014) Synthetic CpG islands reveal DNA sequence determinants of chromatin structure. *Elife*, **3**, e03397.
66. Farcas,A.M., Blackledge,N.P., Sudbery,I., Long,H.K., McGouran,J.F., Rose,N.R., Lee,S., Sims,D., Cerase,A., Sheahan,T.W. *et al.* (2012) KDM2B links the Polycomb Repressive Complex 1 (PRC1) to recognition of CpG islands. *Elife*, **1**, e00205.
67. Wu,X., Johansen,J.V. and Helin,K. (2013) Fbxl10/Kdm2b recruits polycomb repressive complex 1 to CpG islands and regulates H2A ubiquitylation. *Mol. Cell*, **49**, 1134–1146.
68. Kalb,R., Latwiel,S., Baymaz,H.I., Jansen,P.W., Müller,C.W., Vermeulen,M. and Müller,J. (2014) Histone H2A monoubiquitination promotes histone H3 methylation in Polycomb repression. *Nat. Struct. Mol. Biol.*, **21**, 569–571.
69. Blackledge,N.P., Farcas,A.M., Kondo,T., King,H.W., McGouran,J.F., Hanssen,L.L., Ito,S., Cooper,S., Kondo,K., Koseki,Y. *et al.* (2014) Variant PRC1 complex-dependent H2A ubiquitylation drives PRC2 recruitment and polycomb domain formation. *Cell*, **157**, 1445–1459.
70. Jia,D., Jurkowska,R.Z., Zhang,X., Jeltsch,A. and Cheng,X. (2007) Structure of Dnmt3a bound to Dnmt3L suggests a model for de novo DNA methylation. *Nature*, **449**, 248–251.
71. Eggermann,T., Netchine,I., Temple,I.K., Tümer,Z., Monk,D., Mackay,D., Grønskov,K., Riccio,A., Linglart,A. and Maher,E.R. (2015) Congenital imprinting disorders: EUCID.net – a network to decipher their aetiology and to improve the diagnostic and clinical care. *Clin. Epigenet.*, **7**, 23.